\documentclass[12pt]{article}
\usepackage{fullpage}
\usepackage{graphicx}
\usepackage{bm}
\usepackage{amsmath,amssymb,array}
\usepackage{cite}
\usepackage{algorithm,algorithmic}
\DeclareMathOperator*{\argmin}{argmin}
\usepackage{url}

\newcommand{\footremember}[2]{%
	\footnote{#2}
	\newcounter{#1}
	\setcounter{#1}{\value{footnote}}%
}
\newcommand{\footrecall}[1]{%
	\footnotemark[\value{#1}]%
} 

\begin{document}

\title{A New Recurrent Plug-and-Play Prior Based on the Multiple Self-Similarity Network}

\author{
	Guangxiao Song\footremember{1}{Department of Computer Science \& Enginnering, Washington University in St. Louis, MO 63130, USA}\footremember{3}{College of Information Science \& Technology, Donghua University, Shanghai 201620, P.R. China}
	\and
	Yu Sun\footrecall{1}
	\and
	Jiaming Liu\footremember{2}{Department of Electrical \& Systems Engineering, Washington University in St. Louis, MO 63130, USA}
	\and
	Zhijie Wang\footrecall{3}
	\and
	Ulugbek S. Kamilov\footrecall{1} \footrecall{2}
}

\date{}	

\maketitle

\begin{abstract}
Recent work has shown the effectiveness of the plug-and-play priors (PnP) framework for regularized image reconstruction. However, the performance of PnP depends on the quality of the denoisers used as priors. In this letter, we design a novel PnP denoising prior, called multiple self-similarity net (MSSN), based on the recurrent neural network (RNN) with self-similarity matching using multi-head attention mechanism. Unlike traditional neural net denoisers, MSSN exploits different types of relationships among non-local and repeating features to remove the noise in the input image. We numerically evaluate the performance of MSSN as a module within PnP for solving magnetic resonance (MR) image reconstruction. Experimental results show the stable convergence and excellent performance of MSSN for reconstructing images from highly compressive Fourier measurements. 

\end{abstract}


\section{Introduction}
Model-based image reconstruction is often formulated as the following optimization problem
\begin{equation}
	\argmin\limits_{\bm{x}} f(\bm{x}) 
	\;\;\;\;\; \text{with} \;\;\;\;\;
	f(\bm{x})=g(\bm{x})+h(\bm{x})
\label{fomula_1}
\end{equation}
where $\bm{x}$ is the unknown image,
$g$ is the data-fidelity term that penalizes the mismatch to the measurements, and $h$ is the regularizer that imposes a prior knowledge on the unknown.
Over the years, many different regularizers have been proposed for the reconstruction task, including nonnegativity, transform-domain sparsity, and self-similarity \cite{rudin1992nonlinear,figueiredo2001wavelet,elad2006image,danielyan2011bm3d}.
Additionally, a variety of \emph{proximal algorithms} \cite{parikh2014proximal} have been developed to deal with the large amount of data and nondifferentiable regularizers.
Two popular such algorithms are \emph{accelerated proximal gradient method (APGM)} \cite{figueiredo2003algorithm,daubechies2004iterative,bect20041,beck2009fast} and \emph{alternating direction method of multipliers (ADMM)} \cite{eckstein1992douglas,afonso2010fast,ng2010solving,boyd2011distributed}.

Recently, Venkatakrishnan \emph{et al.} \cite{venkatakrishnan2013plug} introduced a powerful \emph{plug-and-play priors (PnP)} framework.
They leveraged the mathematical equivalence of the proximal operator to denoising to infuse state-of-the-art image denoisers, such as BM3D \cite{dabov2007bm3d}, WNNM \cite{gu2014weighted}, TNRD \cite{chen2016trainable}, or DnCNN \cite{zhang2017beyond}, into iterative algorithms.
Although, PnP does not generally minimize any explicit objective function $f$, its effectiveness has been successfully validated on multiple imaging inverse problems \cite{sreehari2016plug,chan2016plug,brifman2016turning,teodoro2016imagerestore,Zhang_2017_CVPR,Teodoro2017sceneadapted,ono2017primaldual,Meinhardt2017learning,Kamilov2017pnppriors,bottou2008tradeoffs,Kamilov2016opticaltomo,ahmad2019plug,sun2019onlinePnP}.

Since the denoiser is the only source of prior knowledge in PnP,
improving it can significantly boost the performance of the recovery.
Recently, non-local neural network based denoisers were shown to achieve the state-of-the-art performance on several image denoising tasks \cite{liu2018non,zhang2019residual}.
In this letter, we propose a new denoiser called \emph{multiple self-similarity net (MSSN)}, based on a recurrent neural network (RNN) and \emph{multi-head attention (MA)} mechanism, to enhance the joint processing of non-local features.
We validate our MSSN denoiser by infusing it into PnP-APGM for magnetic resonance (MR) image reconstruction.
We show that PnP equipped with MSSN outperforms several state-of-the-art denoiser priors by enabling imaging from highly compressive measurements.

\begin{figure*}[htbp]
	\centering
	\includegraphics[width=\columnwidth]{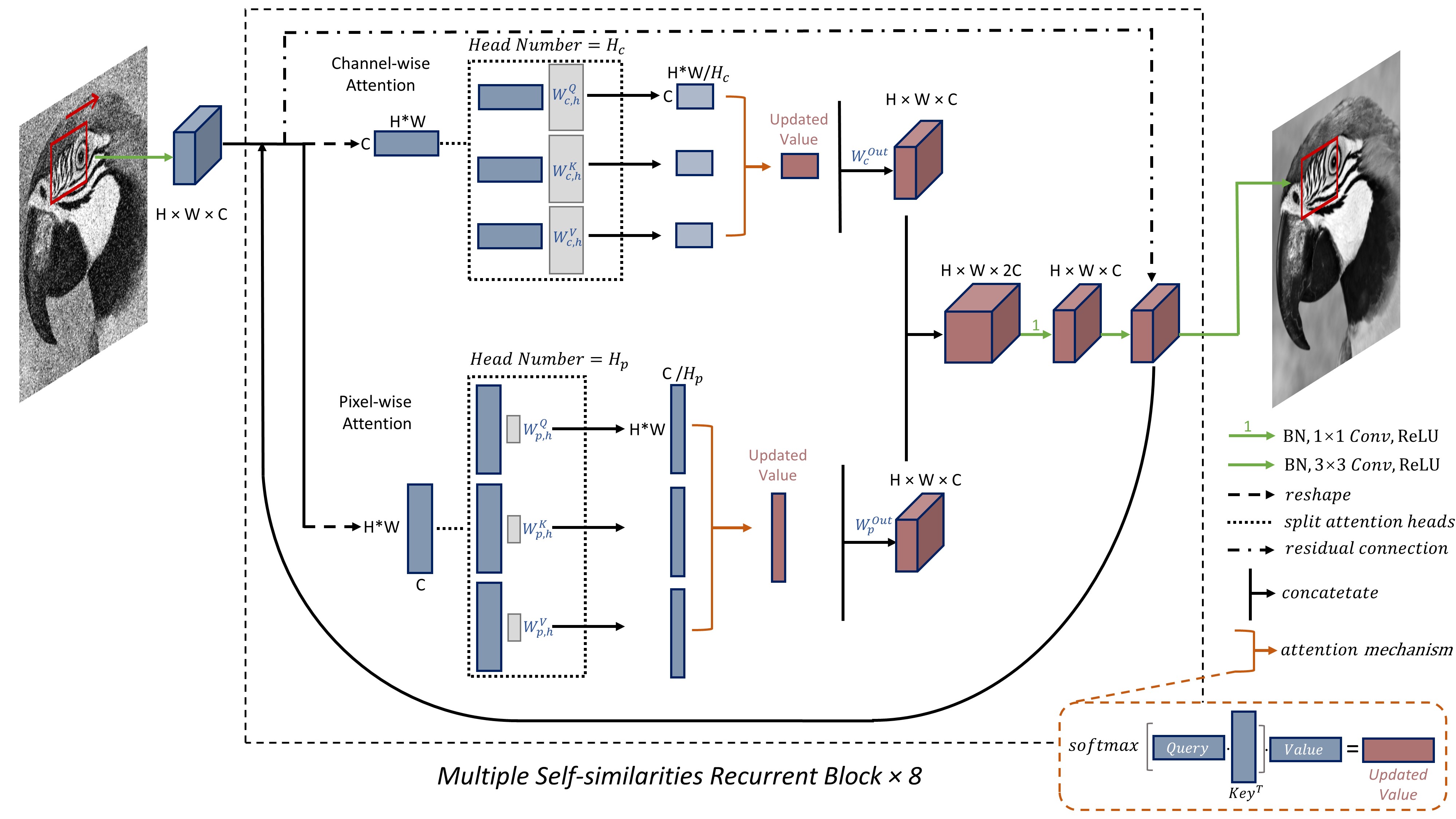}
	\caption{
		An illustration of the proposed MSSN denoiser. Each input patch is obtained from the noisy image stride by stride. After the first convolutional layer, the features are subsequently processed with 8 recurrent blocks. Each recurrent block has dedicated \emph{channel-wise} and \emph{pixel-wise} attention layers consisting of multiple \emph{heads} in order to capture different types of complex relationships. Then two convolutional layers are used for mixing updated features. The final denoised patch is obtained by another convolutional layer.
	}
	\label{fig1_multihead_attention}
\end{figure*}

\section{Background}
In order to solve \eqref{fomula_1} in the context of imaging inverse problems, APGM is widely adopted to avoid the differentiation of the non-smooth $h$.
Recently, Kamilov \emph{et al.} \cite{Kamilov2017pnppriors} proposed a PnP algorithm based on APGM by replacing the proximal operator with a general denoising function. Algorithm  \ref{APGM} summarizes the PnP-APGM, where $\nabla g$ is the gradient of $g$, $\gamma$ is the step size, and sequence $\{q_{k}\}$ are used to accelerate PGM \cite{beck2009fast}. The key advantage of PnP-APGM is that it regularizes the reconstruction by using advanced off-the-shelf denoisers. Similar PnP algorithms have also been developed using ADMM \cite{dong2018denoising}, approximate message passing (AMP) \cite{metzler2016denoising}, Newtons method \cite{buzzard2018plug}, and online processing \cite{sun2019onlinePnP}.
Denoiser priors have also been used within an alternative framework called \emph{regularization by denoising (RED)} \cite{romano2017little,reehorst2018regularization, sun2019block,mataev2019deepRED}.

Different denoisers have been adopted within PnP including BM3D \cite{venkatakrishnan2013plug, Heide2014FlexISP, dar2016postprocessing, rond2016poisson, sreehari2016plug, chan2016plug, Kamilov2017pnppriors, ono2017primaldual, wang2017parameter, sun2019regularized, sun2019onlinePnP}, sparse representations \cite{brifman2016turning}, non-local means (NLM) \cite{sreehari2016plug, unni2018linearized, chan2019performance}, Gaussian mixture models \cite{teodoro2016imagerestore, Teodoro2017sceneadapted, shi2018plug}, and deep-learning-based denoisers.
Among the latter, DnCNN \cite{zhang2017beyond} with residual learning, FFDNet \cite{zhang2018FFDNet} with noise adaptability, and generative adversarial networks (GANs) \cite{goodfellow2014generative} have been particularly popular within PnP \cite{Meinhardt2017learning, ye2018deep, tirer2018image, ryu2019plug, zhang2019deep}.

This letter is based on a hypothesis that further improvements in PnP can be achieved by using \emph{learned non-local priors}.
In conventional denoiers, such as NLM and BM3D, non-local information is infused by jointly processing multiple image patches that are similar to a given reference patch.
In deep neural networks, non-local similarities can be calculated using \emph{attention mechanisms} as was recently done for image restoration in \cite{liu2018non} and \cite{zhang2019residual}.
However, this prior work \cite{liu2018non,zhang2019residual} restricts the number of attention mechanisms to one, which limits the ability of the network to learn more complex non-local information.
In the proposed MSSN, we use \emph{multi-head attention (MA)} to exploit different types of non-local relationships among the feature maps extracted from image, thus boosting the performance of the prior.

\section{Proposed Denoiser Prior}
\label{sec:proposed denoiser}
As illustrated in Fig. \ref{fig1_multihead_attention}, the proposed denoiser mainly consists of two parts.
The first part is a convolutional layer which convolves a noisy patch to produce 128 feature maps.
The second part consists of 8 recurrent blocks for capturing the non-local relationships among the feature maps.
Each block is composed of one multi-head self-attention layer, two convolutional layers, and a residual connection.
Before each convolutional layer, batch normalization is added.
Each convolutional kernel is $3 \times 3$ and ReLU activations are used across the whole network.
Additionally, our denoiser uses the \emph{residual learning} \cite{he2016deep} technique to predict the noise of the input patch.
\begin{algorithm}
	\caption{PnP-APGM}
	\begin{algorithmic}[1]
		\renewcommand{\algorithmicrequire}{\textbf{Input:}}
		\renewcommand{\algorithmicensure}{\textbf{Output:}}
		\REQUIRE $\bm{x}^{0}=\bm{s}^{0} \in \mathbb{R}^{n},\gamma > 0, \sigma > 0, \{q_{k}\}_{k \in \mathbb{N}}$
		\FOR {$k = 1,2,... $}
		\STATE $\bm{z}^{k} \leftarrow \bm{s}^{k-1} - \gamma \nabla g(\bm{s}^{k-1})$
		\STATE $\bm{x}^{k} \leftarrow \text{denoise}_{\sigma}(\bm{z}^{k-1})$
		\STATE $q_{k} \leftarrow \frac{1}{2}(1 + \sqrt{1+4q^{2}_{k-1}})$
		\STATE $\bm{s}^{k} \leftarrow \bm{x}^{k} + ((q_{k-1}-1)/q_{k})(\bm{x}^{k}-\bm{x}^{k-1})$
		\ENDFOR
	\end{algorithmic}
	\label{APGM}
\end{algorithm}
\emph{Multi-head attention (MA)} in our neural network is the core part that captures the complex relationship among the feature maps.
Attention mechanism is widely used for natural language processing (NLP) to estimate the correlation between two sequences.
Similarly, it has the potential to capture the relationships between two feature maps when we regard each row of the feature map as a sequence.
Moreover, MA is able to represent different types of complex relationships while a single attention mechanism learns more simple relationships.
In the proposed denoiser, MA is extended into two types to make the full use of non-local information.
The first is the \emph{pixel-wise attention} and the second is the \emph{channel-wise attention}.

Given a latent representation in terms of keys and values $\bm{z}=[\bm{z}_{1},\bm{z}_{2},...,\bm{z}_{n}]$ , and a query vector $\bm{q}$,
attention calculates the alignment or similarity score between $\bm{q}$ and $\bm{z}_{i}$ (position $i$ of keys) through a function $\theta(\bm{q},\bm{z}_{i})$.
The softmax function normalizes all of the scores $[\theta(\bm{q}, \bm{z}_{i})]_{i=1}^{n}$ to a probability distribution as
\begin{equation}
\textsf{softmax}(\theta(\bm{q}, \bm{z}))_{i} = \frac{ e^{\theta(\bm{q}, \bm{z}_{i})} } {\sum_{j=1}^{n} e^{\theta(\bm{q}, \bm{z}_{j})} } \;,\;\;\;\; i=1,2,...,n.
\end{equation}
The updated value of $\bm{q}$ is calculated as $\sum_{i=1}^{n}  \textsf{softmax}(\theta(\bm{q},\bm{z})_{i})\bm{z}_{i}$.
Dot-product $\theta(\bm{q}, \bm{z}_{i}) = \bm{q}^{\mathsf{T}}\bm{z}_{i}$ is widely used as the function $\theta$.
Note that the attention mechanism makes it easier to learn long range dependencies due to the direct comparison between $\bm{z}_{i}$ and $\bm{q}$.
This $\mathcal{O}(1)$ path length of the attention mechanism makes it preferable to the $\mathcal{O}(n)$ path length of using traditional RNN processing.
Additionally, for self-attention $\bm{q}$ is replaced by $\bm{z}_{p}$ to compute the weight distribution between different positions of latent representations $\bm{z}_{p}$ and $\bm{z}_{i}$. In 2D scenario, the formula of self-attention becomes
\begin{equation}
\label{attention_scores}
\textsf{Attention}(\textsf{Query},\textsf{Key},\textsf{Value})=\textsf{softmax}(\textsf{Query} \textsf{Key}^{\mathsf{T}})\textsf{Value},
\end{equation}
where $\textsf{Query}$, $\textsf{Key}$ and $\textsf{Value}$ are all latent representations $\bm{z} \in \mathbb{R}^{m \times n}$.

Furthermore, multi-head attention is used to capture different types of dependencies between sequences.
\begin{equation}
\begin{aligned}
\label{multihead_attention}
\textsf{MultiHeadAttention}(\textsf{Query},\textsf{Key},\textsf{Value})
=[\textsf{head}_{1},...,\textsf{head}_{h}]W^{Out},
\end{aligned}
\end{equation}
where $\textsf{head}_{h}=\textsf{Attention}(\textsf{Query}W^{Q}_{h},\textsf{Key}W^{K}_{h},\textsf{Value}W^{V}_{h})$, matrices $W^{Q}_{h} \in \mathbb{R}^{n \times d_{k}}$, $W^{K}_{h}  \in \mathbb{R}^{n \times d_{k}}$ and $W^{V}_{h}  \in \mathbb{R}^{n \times d_{v}}$ map \textsf{Query}, \textsf{Key} and \textsf{Value} into lower dimensional spaces $h$ times, to reduce computational cost when we create several attention heads.
After applying \textsf{Attention}, we concatenate the ouputs and pipe through a linear layer $W^{Out} \in \mathbb{R}^{ hd_{v} \times n}$.
In this way, the model can jointly process information from different representation sub-spaces at different positions which improves the capability of matching complex relationships among features.

Our MSSN denoiser adopts this mechanism to exploit potential dependencies in an image.
Specifically, we divide the attention mechanism into \emph{pixel-wise} and \emph{channel-wise} attention layers.
In channel-wise attention, queries, keys, and values are generated by $W^{Q}_{c,h} \in \mathbb{R}^{n \times d_{k}}$, $W^{K}_{c,h}  \in \mathbb{R}^{n \times d_{k}}$ and $W^{V}_{c,h}  \in \mathbb{R}^{n \times d_{v}}$ from the reshaped feature maps $\bm{z} \in \mathbb{R}^{C \times HW}$ along the spatial axis in each attention head.
The computation of matrix multiplication in $\textsf{Attention}(\textsf{Query},\textsf{Key},\textsf{Value})$ function simply compares the similarity between $\textsf{Query}$ and $\textsf{Key}$ in the same spatial position but different channels.
Then the updated values are obtained as in eq. (\ref{attention_scores}).
Similarly, in pixel-wise attention, queries, keys, and values are generated by $W^{Q}_{p,h} \in \mathbb{R}^{n \times d_{k}}$, $W^{K}_{p,h}  \in \mathbb{R}^{n \times d_{k}}$ and $W^{V}_{p,h}  \in \mathbb{R}^{n \times d_{v}}$ from $\bm{z} \in \mathbb{R}^{HW \times C}$ along the channel axis in each attention head.
Similarities of feature maps in the same channel, but in different spatial positions are calculated.
Lastly, we concatenate the updated pixel-wise and channel-wise values obtained by $W_{p}^{Out} \in \mathbb{R}^{ hd_{v} \times n}$ and $W_{c}^{Out} \in \mathbb{R}^{ hd_{v} \times n}$,
then dimension reduction is done using another $1 \times 1 $ convolutional layer.
The mixed attention values are obtained in the same shape of the input feature maps.
\begin{table}[tp!]
	\centering
	\caption{
		Average SNRs of MR Image Reconstruction for Randomly Selected Samples
	}
	\begin{tabular}{|c|c|c|c|c|c|c|c|}
		\hline
		& IFFT   & TV     & BM3D   & DnCNN & DnCNN (MRI) & SSN    & MSSN             \\ \hline
		\textit{36 Lines} & 17.36 & 20.03 & 20.62 & 19.15 & 20.17 & 21.01 & \textbf{21.18} \\ \hline
		\textit{48 Lines} & 18.54 & 21.27 & 21.84  & 20.58 & 21.37 & 21.71 & \textbf{22.03} \\ \hline
	\end{tabular}
	\label{table1_results}
\end{table}
\section{Experiments}
We validate the proposed denoiser within PnP-APGM for MRI reconstruction by comparing it with three popular priors: total variation (TV) \cite{rudin1992nonlinear}, BM3D \cite{dabov2007bm3d}, and DnCNN \cite{zhang2017beyond}.
Specifically, we reconstruct from the subsampled $k$-space measurements $\bm{y}$ and the data-fidelity term in eq. (\ref{fomula_1}) becomes 
\begin{equation}
g(\bm{x}) = \frac{1}{2} \lVert \bm{y} - \mathcal{S} \circ \mathcal{F}(\bm{x}) \rVert_{2}^{2} \;,
\end{equation}
where $\mathcal{F}$ denotes the Fourier transform, $\mathcal{S}$ is subsampling, and $\circ$ denotes the composition. 
We use 10 randomly-picked samples\footnote{Data no.~58, 136, 200, 4008, 13196, 15471, 17000, 21119, 27900, 31135} from the open-source fastMRI dataset \cite{zbontar2018fastMRI} for testing.
\subsection{Experimental Settings}
To test the robustness of our MSSN denoiser as a PnP prior,
we trained it on natural images from the standard BSD500 dataset, but tested it on MRI images.
MSSN is trained on patches of size $42 \times 42$ using the Gaussian noise with $\sigma = 5$.
The number of recurrent blocks is 8 and the number of features for each $3 \times 3$ convolutional kernel is set to 128 across all the network.
The number of heads for both types of attention (\emph{pixel-wise} and \emph{channel-wise}) are set to 2.
Adam algorithm is adopted to optimize the neural network by minimizing the \emph{mean square error (MSE)} loss.
The learning rate starts at $10^{-3}$, then decreases by half every $5 \times 10^{4}$ iterations with the epoch number of 80.

Radial sampling method in $k$-space is used for acquisition. 
We used highly compressive 36 and 48 lines in our experiments.
We optimized the hyperparameters of TV and BM3D for SNR.
We provided comparisons with two variants of DnCNN, one trained on the same set of natural images as MSSN and the other trained specifically on the fastMRI dataset (labeled as DnCNN (MRI)).
Both DnCNN denoisers were trained with noise $\sigma = 5$
For better comparison between the proposed mixed multi-head attention and single attention in the RNN denoiser, these two attention mechanisms are tested (referred as MSSN and SSN, respectively).

\subsection{Results and Discussions}
The quantitative results in Table \ref{table1_results} show that MSSN outperforms the optimized conventional regularizer TV, BM3D, and DnCNN denoisers.
MSSN achieves the highest average SNR of 21.18 dB when using 36-lines $k$-space subsampling and 22.03 dB when using 48 lines.
The SSN denoiser is competitive with BM3D without using the mixed multi-head attention.
The MSSN denoiser further improves the reconstruction performance compared to SSN, and results in the lowest reconstruction error, even when trained on natural images.

\begin{figure*}[htbp]
	\includegraphics[width=\columnwidth]{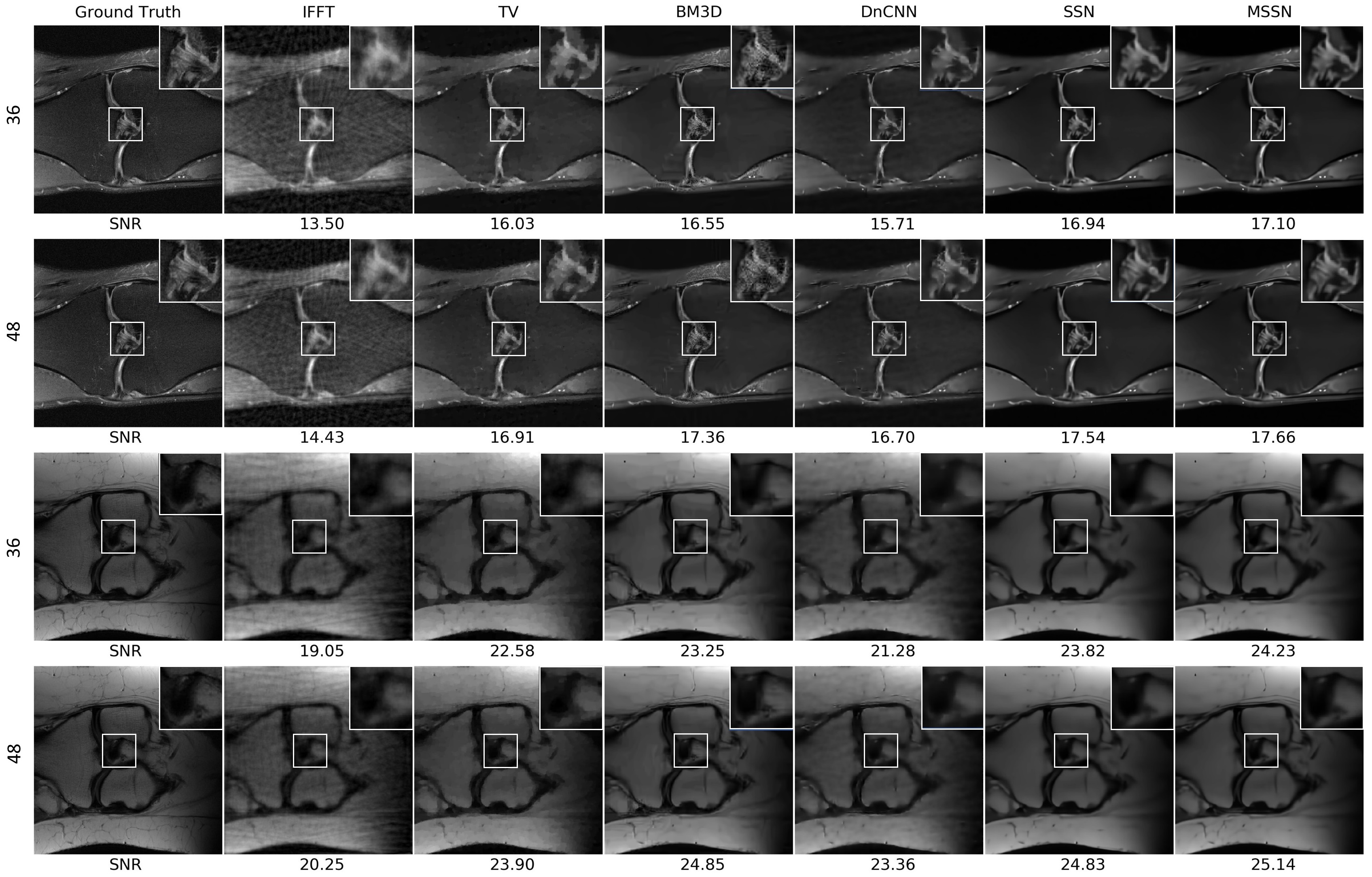}
	\caption{
		Illustrations of MRI reconstruction from $k$-space downsampled data (\emph{no.~58} and \emph{no.~21119} using radial sampling, the first and third rows use 36 lines for sampling and the second and the fourth rows use 48 lines) using different denoisers within PnP.
		Each reconstruction is labeled with its SNR (dB) value with respect to the ground truth.
	}
	\label{fig2_results}
\end{figure*}

Figure \ref{fig2_results} demonstrates the details of the MRI reconstruction.
The first two rows are the reconstruction results of \emph{no.~58} in fastMRI dataset using 36-lines and 48-lines radial sampling respectively.
The last two rows are results of \emph{no.~21119}. 
Visual inspection of reconstructed images reveals that proposed denoiser is robust in these two T1, T2-weighted MR images of bones, and artefacts are diminished using SSN or MSSN denoiser compared to conventional denoisers.
More crucially, PnP algorithm using MSSN keeps subtle yet important anatomical details and offers more faithful reconstructions that can be recognized in the zoomed-in views.
\section{Conclusion}
In this letter, we proposed and validated a new PnP denoiser called MSSN, bringing together two concepts, namely RNN and MA.
MSSN captures multiple self-similarities of non-local features.
We combined MSSN with PnP-APGM for solving an inverse problem in MRI.
Experimental results on fastMRI dataset show that MSSN improves both quantitative and perceptual quality, even when trained on natural images.
This suggests that the powerful feature extraction of deep RNN and the ability to capture complex non-local relationships of the MA mechanism (in both pixel-wise and channel-wise attention) makes MSSN competitive with the state-of-the-art PnP priors.


\end{document}